\documentclass[usenatbib]{mnras}

\usepackage[T1]{fontenc}
\usepackage{ae,aecompl}

\usepackage{graphicx}	
\usepackage{amssymb}
\usepackage[fleqn]{amsmath}

\DeclareMathOperator\erfc{erfc}

\newcommand{\kms}{km\,s$^{-1}$}
\newcommand{\mjykmsbm}{mJy\,km\,s$^{-1}$\,beam$^{-1}$}
\newcommand{\mjykms}{mJy\,km\,s$^{-1}$}
\newcommand{\intlflux}{9.5$\pm$3.6~mJy\,\kms}
\newcommand{\pvel}{8.5$\pm$0.2~\kms}
\newcommand{\comass}{(1.4--13)$\times10^{-6}$~M$_\oplus$}
\begin{document}

\title[Deep ALMA Search for CO Gas in the HD 95086 Debris Disc]{Deep ALMA Search for CO Gas in the HD 95086 Debris Disc}
\author[M. Booth \& co.]{Mark Booth$^{1}$\thanks{E-mail: markbooth@cantab.net}, Luca Matr\`a$^{2}$, Kate Y. L. Su$^{3,4}$, Quentin Kral$^{5,6}$, \newauthor Antonio S. Hales$^{7,8}$, William R. F. Dent$^{7}$, A. Meredith Hughes$^{9}$, \newauthor Meredith A. MacGregor$^{10}$, Torsten L\"ohne$^{1}$ and David J. Wilner$^{2}$ \\
$^{1}$ Astrophysikalisches Institut und Universit\"atssternwarte, Friedrich-Schiller-Universit\"at Jena, Schillerg\"a\ss{}chen 2-3, 07745 Jena, \\Germany \\
$^{2}$ Harvard-Smithsonian Center for Astrophysics, 60 Garden Street, Cambridge, MA 02138, USA \\
$^{3}$ Steward Observatory, University of Arizona, 933 N Cherry Avenue, Tucson, AZ 85721, USA \\
$^{4}$ Institute of Astronomy and Astrophysics, Academia Sinica, P.O. Box 23-141, Taipei 106, Taiwan \\
$^{5}$ Institute of Astronomy, University of Cambridge, Madingley Road, Cambridge CB3 0HA, UK \\
$^{6}$ LESIA, Observatoire de Paris, Universit\'e PSL, CNRS, Sorbonne Universit\'e, Univ. Paris Diderot, Sorbonne Paris Cit\'e, 5 place \\ Jules Janssen, 92195 Meudon, France \\
$^{7}$ Joint ALMA Observatory, Alonso de C\'ordova 3107, Vitacura 763-0355, Santiago, Chile \\
$^{8}$ National Radio Astronomy Observatory, 520 Edgemont Road, Charlottesville, Virginia, 22903-2475, USA \\
$^{9}$ Department of Astronomy, Van Vleck Observatory, Wesleyan University, Middletown, CT 06459, USA \\
$^{10}$ Department of Terrestrial Magnetism, Carnegie Institution for Science, 5241 Broad Branch Road, Washington, DC 20015, USA
}

\date{Accepted 2018 October 31. Received 2018 October 30; in original form 2018 July 18}

\maketitle

\begin{abstract}
One of the defining properties of debris discs compared to protoplanetary discs used to be their lack of gas, yet small amounts of gas have been found around an increasing number of debris discs in recent years. These debris discs found to have gas tend to be both young and bright. In this paper we conduct a deep search for CO gas in the system HD~95086 -- a 17~Myr old, known planet host that also has a debris disc with a high fractional luminosity of $1.5\times10^{-3}$. Using the Atacama Large Millimeter/submillimeter Array (ALMA) we search for CO emission lines in bands 3, 6 and 7. By implementing a spectro-spatial filtering technique, we find tentative evidence for CO $J$=2-1 emission in the disc located at a velocity, \pvel, consistent with the radial velocity of the star. The tentative detection suggests that the gas on the East side of the disc is moving towards us. In the same region where continuum emission is detected, we find an integrated line flux of \intlflux, corresponding to a CO mass of \comass. Our analysis confirms that the level of gas present in the disc is inconsistent with the presence of primordial gas in the system and is consistent with second generation production through the collisional cascade.
\end{abstract}

\begin{keywords}
circumstellar matter -- planetary systems -- submillimetre: planetary systems -- submillimetre: stars -- stars: individual: HD~95086
\end{keywords}

\section{Introduction}
The presence of gas in a circumstellar disc was once used as a factor in distinguishing between protoplanetary and debris discs. As a system evolves, the primordial gas is depleted by photoevaporation and accretion onto the star and growing planets. The disappearance of the gas allows the remaining planetesimals to be dynamically stirred, increasing their relative velocities and enabling the onset of a collisional cascade. Such collisions produce debris and hence the circumstellar disc becomes a debris disc \citep[see e.g.][]{wyatt15}.

Nonetheless, a number of systems that otherwise appear to be debris discs do show signs of the presence of atomic and molecular gas in both absorption and emission \citep[see][for a recent review]{hughes18}. So far the only molecular gas that has been observed in emission is CO. In some cases the CO gas seems to be consistent with being a primordial remnant from the protoplanetary disc. HD~21997 is the best example of this with a CO mass comparable to its dust mass and with CO gas interior to the cold dust belt despite its $\sim30$~Myr age \mbox{\citep{kospal13,moor13a}}. On the other hand, there are three cases where the CO is clearly colocated with the cold dust belt and the CO density is so low that it must be optically thin. These are HD~181327 \citep{marino16}, $\beta$ Pic \citep{kral16,matra17} and Fomalhaut \citep{matra17a}. When the gas is optically thin to UV light, it cannot shield itself from photodissociating radiation and so the CO molecules are broken down on short timescales \citep{visser09}. Since the gas must then be replenished on timescales much shorter than the age of the system, the gas observed in these cases cannot be of a primordial origin.

HD~95086 is located at a distance of 86.4$\pm$0.2~pc \citep[\emph{Gaia} Data Release 2;][]{gaia18}\footnote{Radial distances quoted from papers using the \emph{Hipparcos} or \emph{Gaia} Data Release 1 distances have been adjusted to the new distance throughout this paper.} and is part of the Scorpius-Centaurus Association \citep[specifically part of the Lower Centaurus Crux;][]{dezeeuw99}. It hosts a 4-5~M$_J$ planet with a semi-major axis, $a=54^{+13}_{-25}$~AU and eccentricity, $e=0.2^{+0.3}_{-0.2}$ \citep{chauvin18}. The debris disc was first resolved by \emph{Herschel} \citep{moor13,su15}. Analysis of recent images from ALMA shows the cold component of the disc to extend from around 100 to 320~AU \citep{su17a,zapata18}. In addition to the disc, two bright, compact sources are found on the West side of the disc and are thought to be background contamination.

Like most of the other systems where gas has been detected, HD~95086 is an A type star, with a luminosity of $L_\star=6.63\pm0.03$ L$_\odot$ \citep{gaia18}, and is young, with an estimated age of 17$\pm$4~Myr \citep{meshkat13}, meaning that it could plausibly still retain its primordial gas disc as could potentially be the case for other stars in Sco-Cen \citep{lieman16,moor17}. However, the non-detection of gas presented by \citet{zapata18} shows that this is not the case. Nonetheless, with its high fractional luminosity of $f=1.5\times10^{-3}$ \citep{rhee07}, the disc around HD~95086 also represents one of the best candidates for containing second generation gas \citep{kral17}. The first attempt to find CO around HD~95086 was undertaken by \citet{moor13} using APEX to search for the $J$=3-2 emission line. They did not detect any CO and placed an upper limit of 1.95~Jy\,\kms. With ALMA observations a sensitivity three orders of magnitude greater is achievable, as demonstrated by \citet{zapata18} who first published limits on the integrated CO $J$=3-2 and $J$=2-1 emission using the data that is also discussed in this paper. Here we use a spectro-spatial filtering technique \citep{matra15, matra17a} to further explore the data in an attempt to detect the very low mass of CO that may be present in this system and do find tentative evidence of its presence.

\section{Observations}
HD~95086 was first observed by two ALMA programmes in cycle 2: \#2013.1.00773.S (PI: Su; hereafter referred to as data set A) and \#2013.1.00612.S (PI: Booth; hereafter referred to as data set B). In both cases, the observations were taken in band 6 whilst the array was in a compact configuration. The total on-source observation time for the combined data sets is 9.1 hours. Further details of the observations along with analysis of the continuum data can be found in \citet{su17a}.

Both data sets cover the CO $J$=2-1 line (which has a rest frequency of 230.538~GHz) with one spectral window, which has 3840 channels of width 0.63~\kms{} or 0.49~MHz. We reduced both data sets using \texttt{casa} version 4.7.2 in pipeline mode. We subtracted the continuum from each data set using \texttt{uvcontsub} by fitting to the parts of the spectral window that do not include the line. We then inverted the visibilities using \texttt{CLEAN} to create the data cube using natural weighting to optimise the signal to noise ratio. The data cube for data set A has an RMS (root mean squared) per spectral channel of 0.50~mJy\,beam$^{-1}$, the data cube for data set B has an RMS per spectral channel of 0.61~mJy\,beam$^{-1}$ and the data cube of the combined data has an RMS per spectral channel of 0.38~mJy\,beam$^{-1}$. The angular resolution is $1.5\times1.3\arcsec{}$ for both data sets.

\begin{figure}
	\centering
	\includegraphics[width=0.48\textwidth]{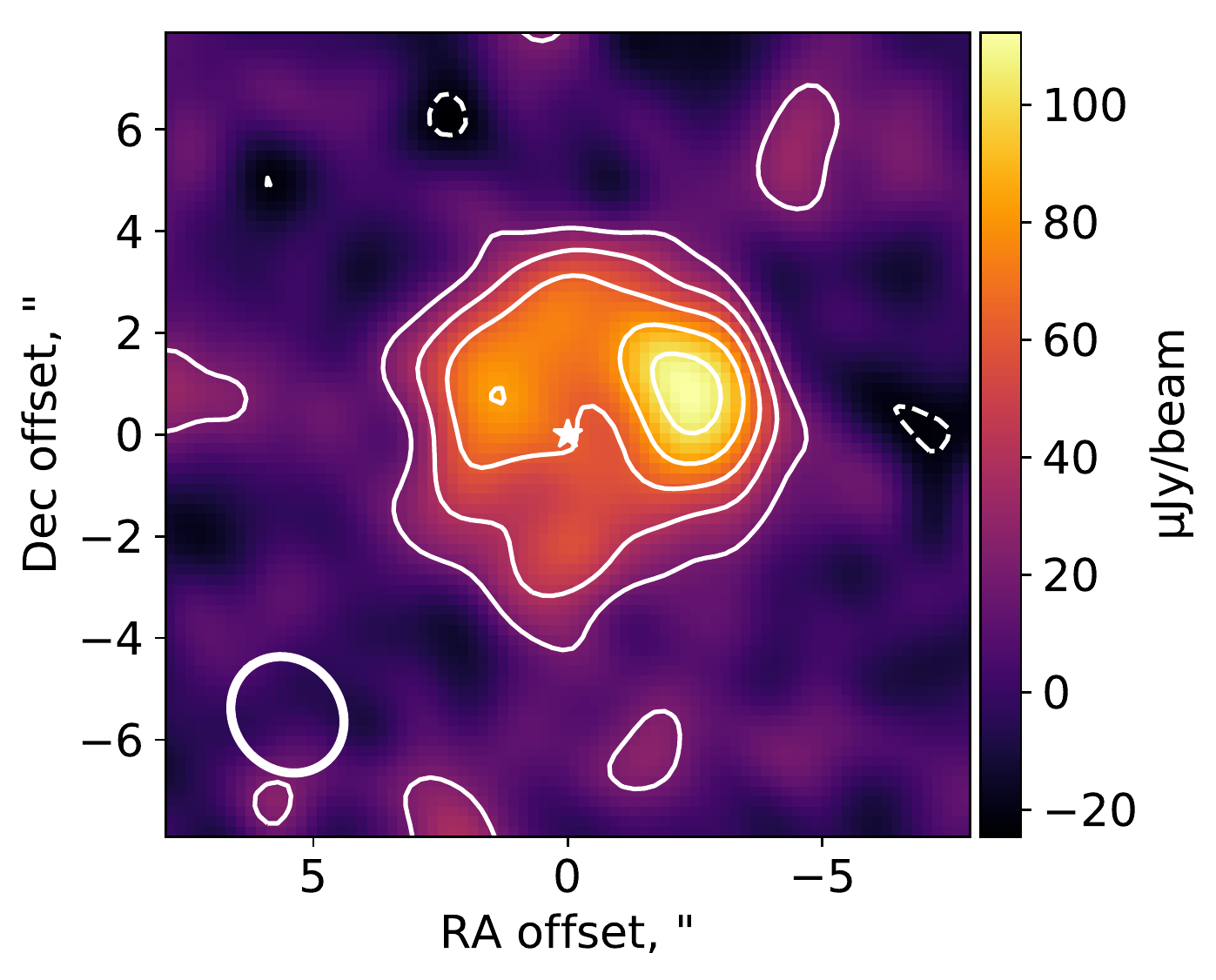}
	\caption{Band 3 (2.8~mm) continuum emission. The image has been created using natural weights and a primary beam correction has been applied. The ellipse in the corner represents the beam size ($2.39\arcsec\times2.11\arcsec$). Contours are in $\pm2\sigma$ increments, where $\sigma=10\,\mu$Jy\,beam$^{-1}$.}
	\label{fb3cont}
\end{figure}

HD~95086 was also observed by ALMA in bands 3 and 7 under a directors discretionary time proposal: \#2016.A.00021.T (PI: Ho). These data sets were also reduced using \texttt{casa} version 4.7.2 in pipeline mode. The band 3 data were for a total on-source observing time of 55 minutes. We present this here for the first time and show the continuum image in figure \ref{fb3cont}, created using \texttt{CLEAN} with natural weights. This observation also included a spectral window covering the CO $J$=1-0 line (which has a rest frequency of 115.271~GHz) with 3840 channels of width 0.63~\kms{} or 0.24~MHz. The band 7 data were for a total on-source observing time of 6 minutes and is discussed in detail in \citet{zapata18}. It included a spectral window covering the CO $J$=3-2 line (which has a rest frequency of 345.796~GHz) with 3840 channels of width 0.42~\kms{} or 0.49~MHz. As with the band 6 data, the continuum was subtracted from each of these and the visibilities were inverted creating a data cube with an RMS per spectral channel of 2.2~mJy\,beam$^{-1}$ and angular resolution of $1.9\times1.7\arcsec{}$ for the band 3 data and an RMS per spectral channel of 4.8~mJy\,beam$^{-1}$ and angular resolution of $0.8\times0.5\arcsec{}$ for the band 7 data. 

\section{CO emission lines}
\subsection{CO $J$=2-1 analysis}
\label{s21}
\begin{figure*}
	\centering
	\begin{tabular}{cc}
	\includegraphics[width=0.48\textwidth]{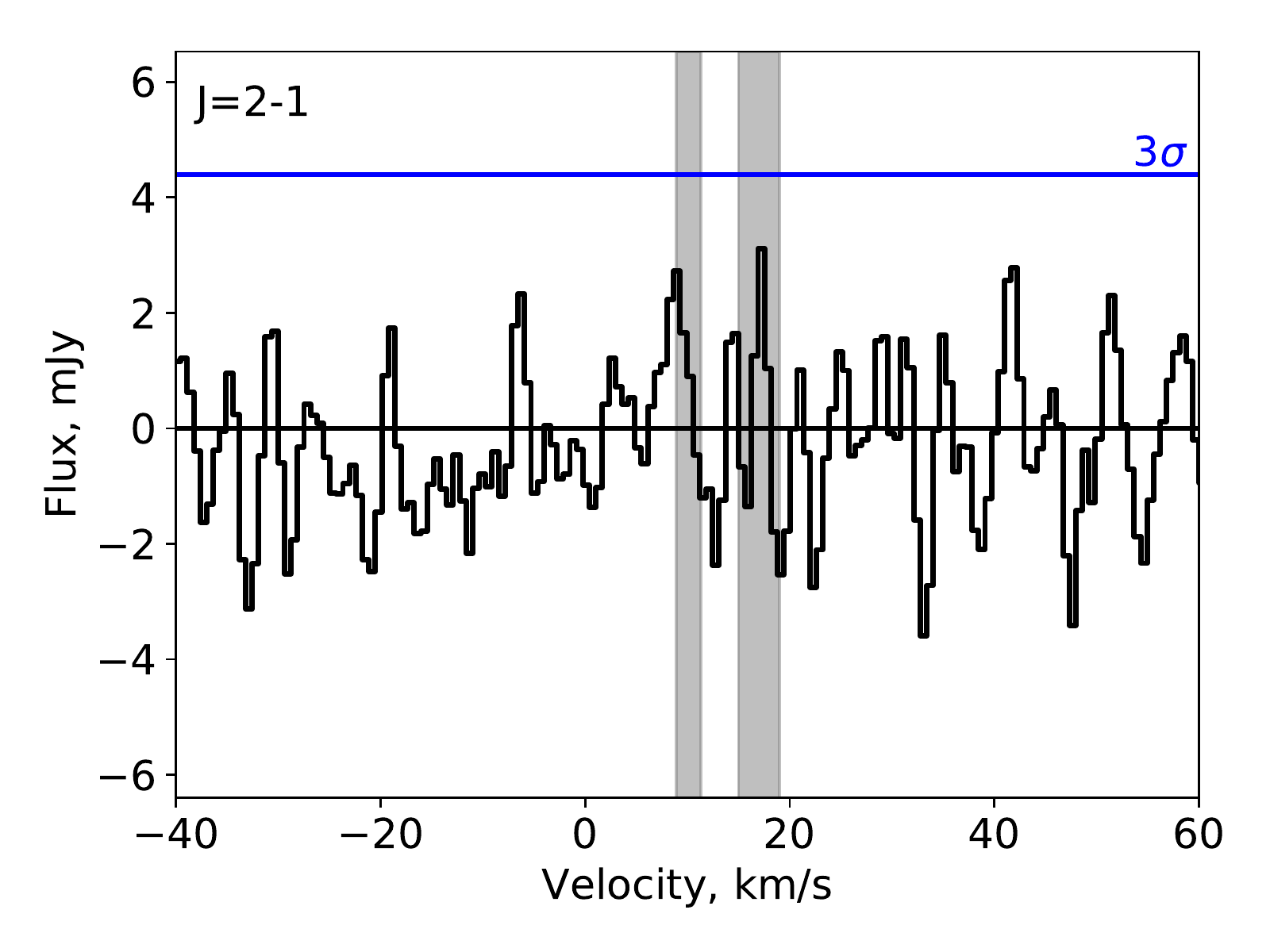} &
	\includegraphics[width=0.48\textwidth]{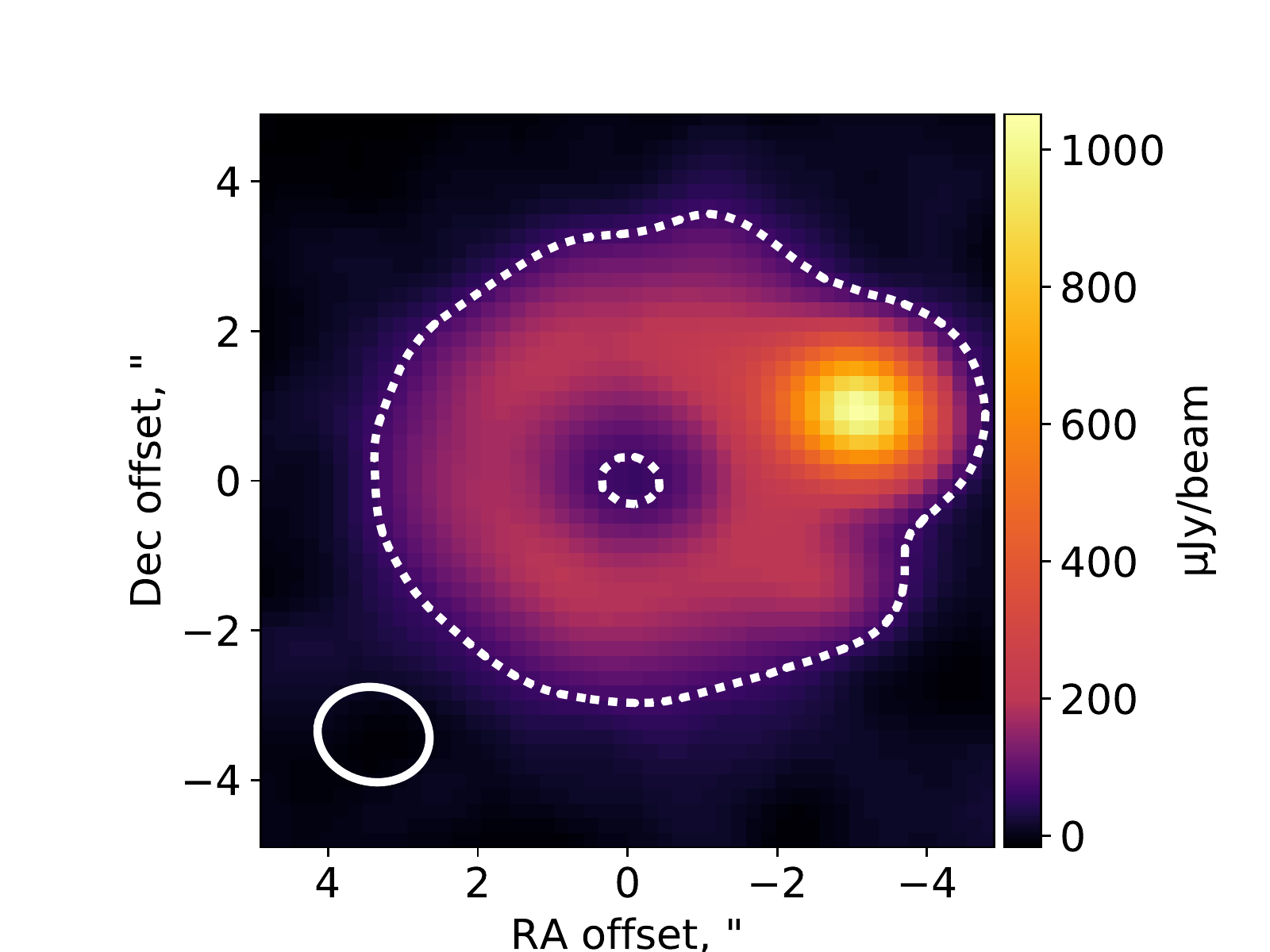} \\
	\includegraphics[width=0.48\textwidth]{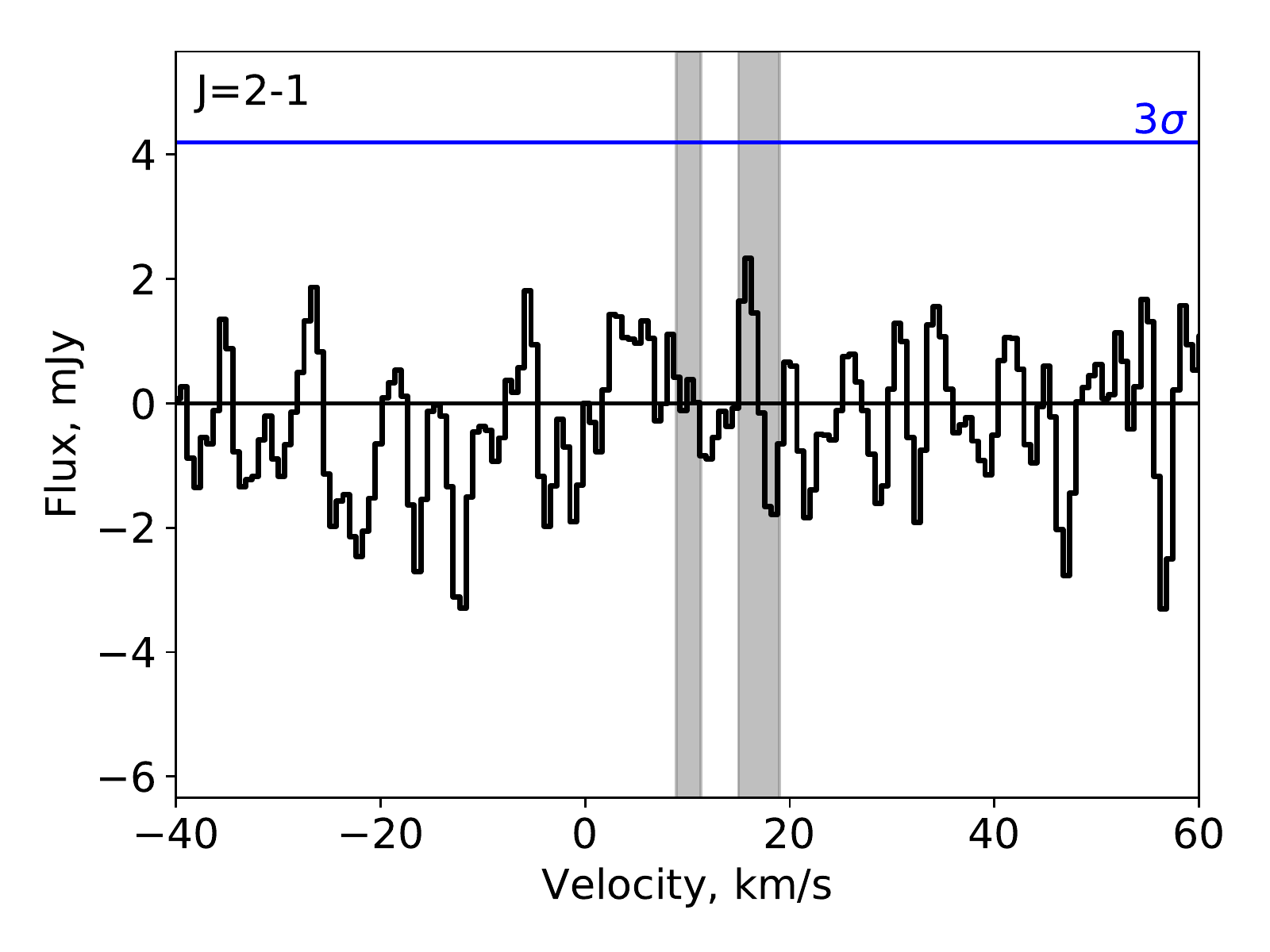} &
	\includegraphics[width=0.4\textwidth]{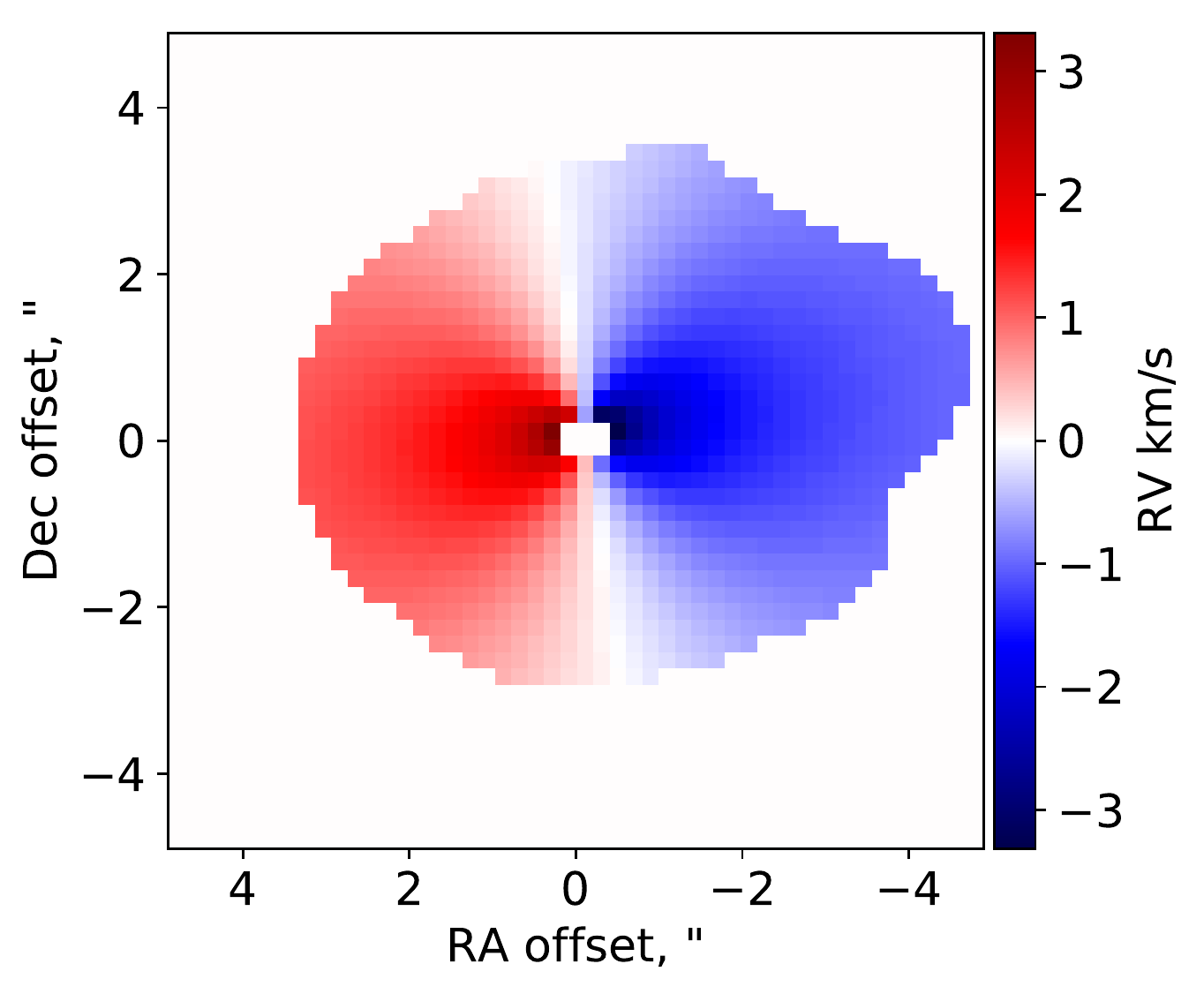} \\
	\includegraphics[width=0.48\textwidth]{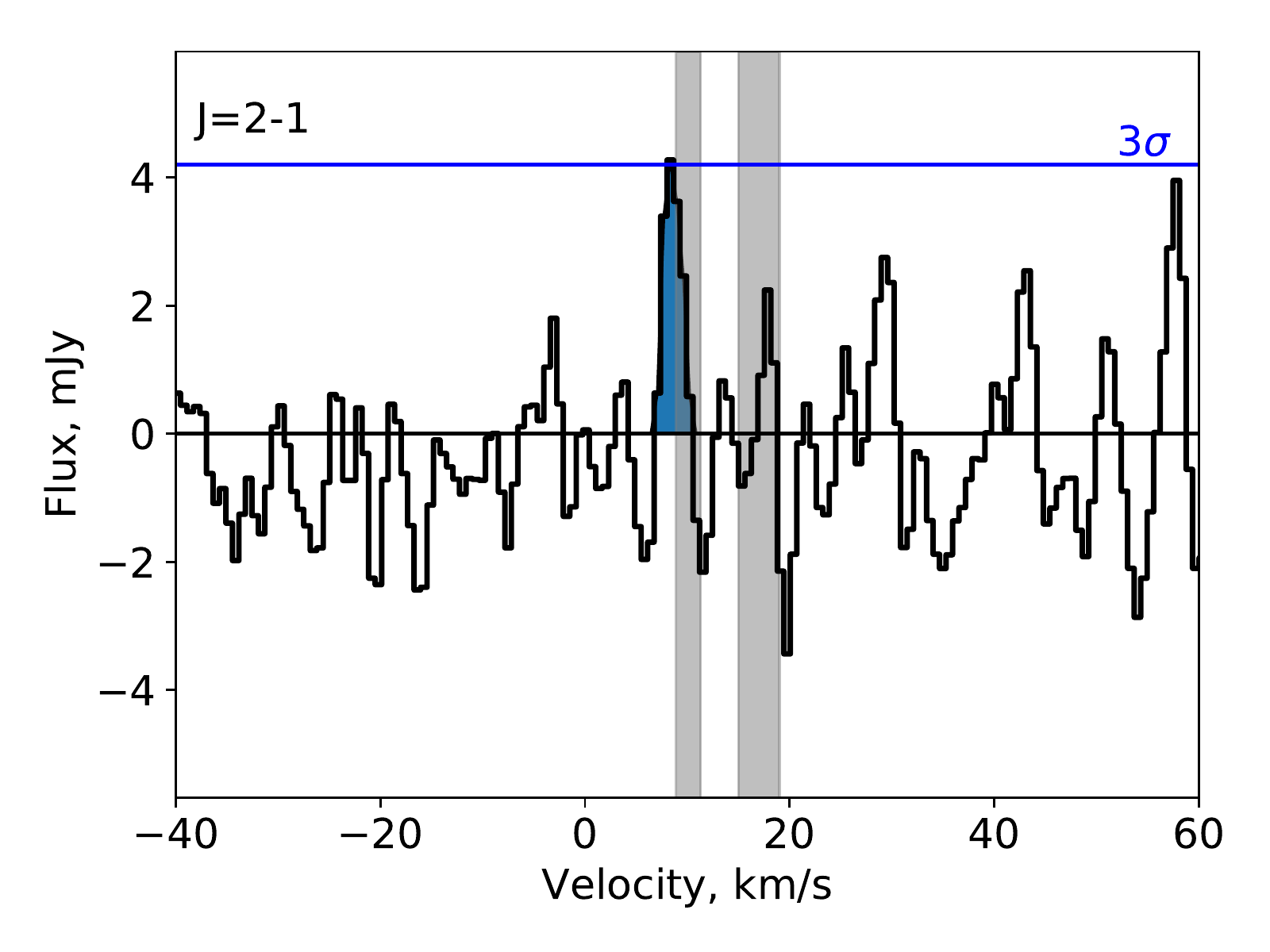} &
	\includegraphics[width=0.4\textwidth]{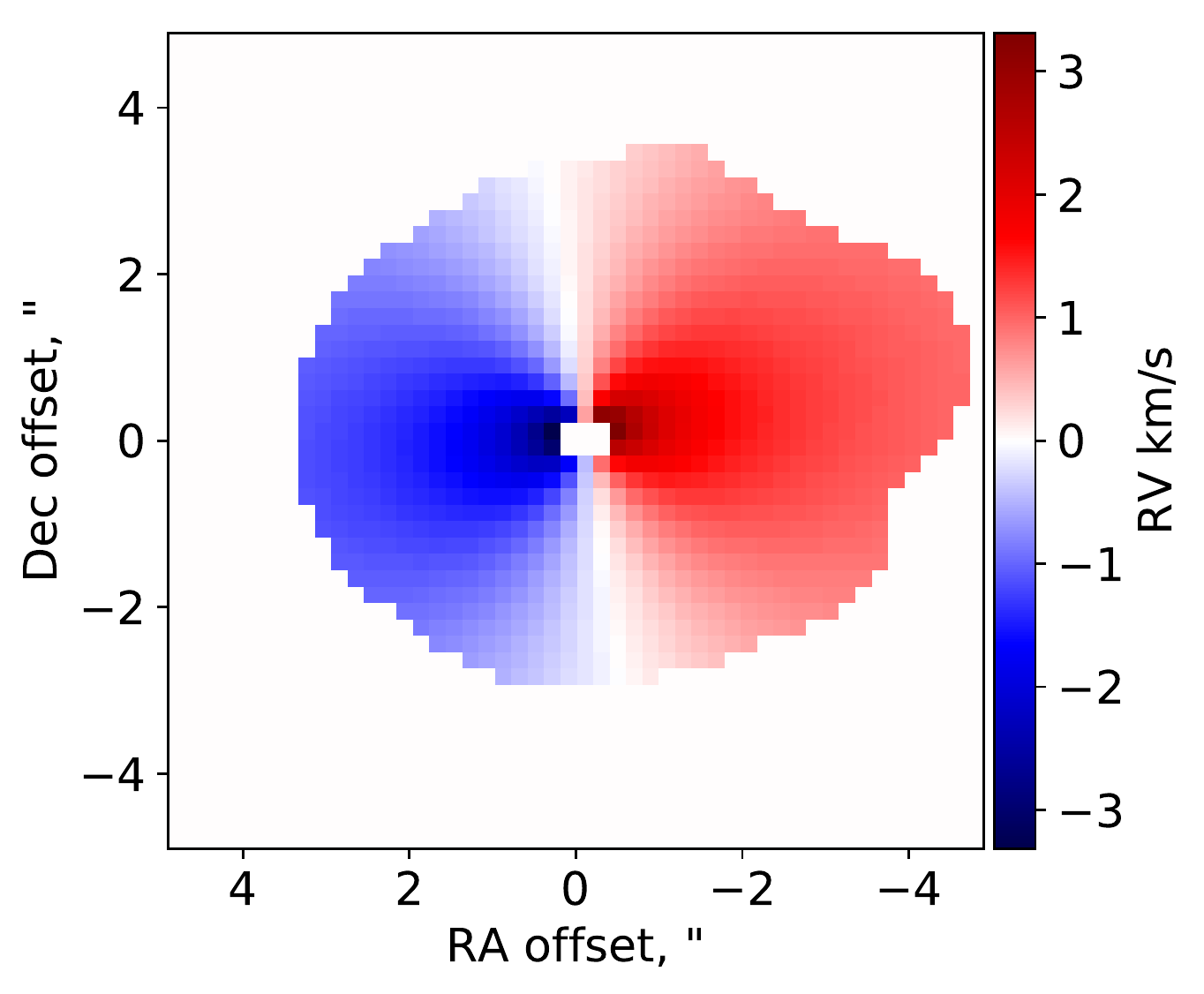} \\
	\end{tabular}
	\caption{The plots on the left show the spectral profile maps spatially integrated over the disc. The velocity reference frame is the barycentric frame for the CO $J$=2-1 line and the horizontal line represents 3$\sigma$ significance. The top plot on the left shows the profile without any spectral shifting and the top plot on the right shows the band 6 CLEAN continuum image with a dashed line indicating the region used for the spatial integration. The middle plot shows the case where we assume the gas to the East is moving away from us and the gas to the West is moving towards us and vice versa for the bottom plot (as shown by the radial velocity plots on the right hand side). The vertical shaded regions show the two measurements and 1$\sigma$ uncertainties for the radial velocity of the star (10.1$\pm$1.2~km\,s$^{-1}$ from \protect\citet{madsen02} and 17$\pm$2 \kms{} from \protect\citet{moor13}). The strength of the peak in the bottom plot is a good indication that this sense of rotation is the correct one.}
	\label{fvel}
\end{figure*}

Although there is no obvious CO detection in the dirty data cube, it is possible that there may still be evidence for CO if we make use of spatial and spectral filters as \citet{matra15,matra17a} have done for Fomalhaut. We start by looking for evidence of CO $J$=2-1 emission in the band 6 combined data as this is the deepest dataset. The top plot of figure \ref{fvel} shows the spectrum obtained by summing over all pixels with a signal-to-noise ratio of at least 10 in the continuum image (this covers radii of $\sim$0.5\arcsec{} to $\sim$5\arcsec{} from the star)\footnote{For this we use the natural weighted continuum image, which has an RMS of 6.7~mJy\,beam$^{-1}$. Note that this is different to the value given in \citet{su17a} and the beam size is also different as they actually used Briggs weighted images with a robust parameter of 0.5 throughout that paper but erroneously referred to them as natural weighted images.}. The radial velocity given is the barycentric radial velocity set such that a CO line with velocity equal to the Solar System's barycentric velocity would appear at 0~\kms. If CO is present in the disc, it should have a velocity consistent with that of the star. In the case of HD~95086, there are two values in the literature for the radial velocity of the star. \citet{moor13} used spectroscopy to determine a radial velocity of 17$\pm$2~\kms{}. Unfortunately, radial velocities are difficult to determine for early-type stars due to their high rotational velocities and because their optical spectra have few absorption features \citep{becker15}. An alternative is to consider the three dimensional space motions of stars. \citet{madsen02} used Hipparcos astrometry to determine a radial velocity of 10.1$\pm$1.2~\kms. Given the discrepant nature of these radial velocities, we assume that  CO from the debris disc will have a velocity close to one of these velocities. These velocities are indicated in the plots by the vertical lines. There is a rise in the spectrum at both velocities -- 2.7$\pm$1.5~mJy at around 9~\kms{} and 3.1$\pm$1.5~mJy at around 17~\kms{} -- but in both cases these peaks have a significance of $<3\sigma$ (where $\sigma$ is the RMS of the spectrum) and so are indistinguishable from the noise. From this we can only conclude that the RMS on the integrated line flux spatially integrated across the disc is 2.4~mJy\,\kms{} (see Appendix \ref{sapp1}).

Due to the orbital motion of the gas and the non-zero inclination of the disc, the gas will likely have a slightly different radial velocity to that of the star, for instance an axisymmetric gas disc will produce a double-peaked spectrum if spectrally resolved. For a low signal-to-noise ratio (SNR) observation such as this, we can attempt to increase the SNR by taking into account the velocity and shifting the spectra in each pixel so that it should peak at the velocity of the star. Assuming the gas is moving at Keplerian velocity \citep[as is shown to be the case for the $\beta$ Pic disc,][]{dent14}, the radial component of the velocity at a point in the disc $(R, \theta)$ is then given by:
\begin{equation}
v_{\rm{rkep}}(R,\theta)=29.8\sqrt{\frac{M_\star/\rm{M}_\odot}{R/\rm{AU}}}\sin(I)\cos(\theta-\Omega)\,\rm{km/s}
\end{equation}
where $M_\star$ is the mass of the star ($\sim$1.6\,M$_\odot$ for HD~95086 assuming a ${M_\star}/{M_\odot}={\sqrt[4]{{L_\star}/{L_\odot}}}$ mass-luminosity relation), $R$ is the distance to the star, $I$ is the inclination, $\theta$ is the angular displacement of the pixel measured anti-clockwise from North and $\Omega$ is the position angle of the disc's semi-major axis. We assume the gas disc has the same inclination and position angle as the dust disc, which are found to be around 31$\degr$ and 98$\degr$ respectively \citep{su17a}. Whether the velocity is positive or negative at any given disc location will depend on which direction the gas is orbiting. The plots on the right in figure \ref{fvel} show the relative radial velocities across the disc assuming the gas is travelling at keplerian velocity for the given orbital distance of each pixel. The middle plot is for the case where the gas to the East is moving away from us and the bottom plot is for the case where the gas to the East is moving towards us. Shifting the spectra in each pixel by the negative of its assigned radial velocity relative to that of the star and then summing spatially across the region where the continuum is detected gives the spectral profiles shown in figure \ref{fvel} middle and bottom for the East side moving away and towards us respectively.

We find that no signal is detected in the case where the East side is moving away from us. On the other hand, for the case where the East side is moving towards us, the previous tentative signal at $\sim$9~\kms{} has been boosted from 2.7$\pm$1.5~mJy to 4.3$\pm$1.4~mJy therefore implying that the gas is likely orbiting in this manner. 

\begin{figure}
	\centering
	\includegraphics[width=0.48\textwidth]{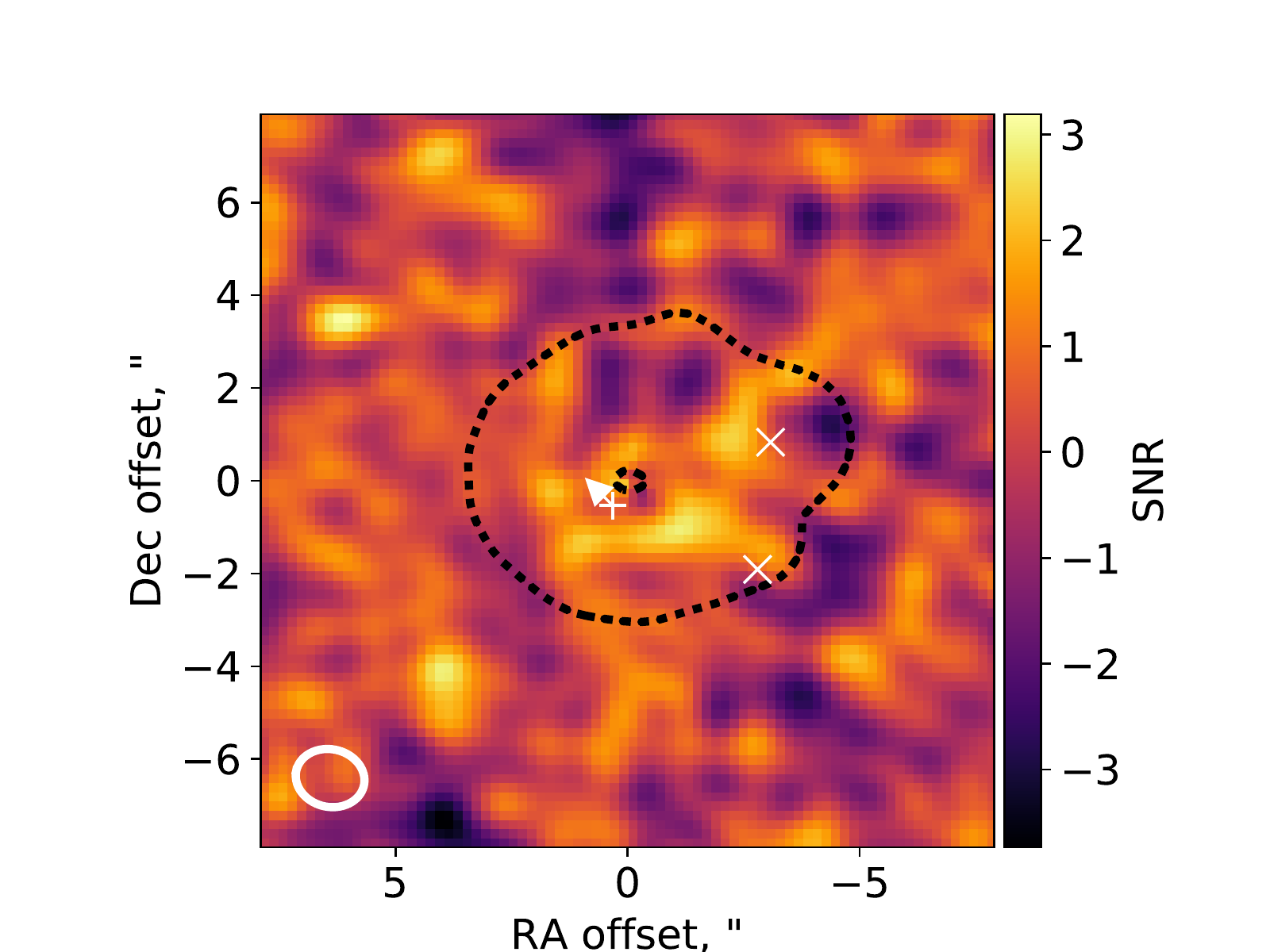}
	\caption{SNR map of the CO $J$=2-1 line after spectral shifting and integrating over the channels that cover the CO $J$=2-1 line (the blue channels in figure \ref{fvel}). The crosses mark the locations of the two bright continuum sources, the upper of which is noticeably brighter than the rest of the disc in continuum emission (see section \ref{sbright} and the top right plot of figure \ref{fvel}). The dashed line marks the region where the continuum emission has a signal-to-noise ratio of at least 10. The plus marks the location of the planet along with the direction of its orbit as observed by the Gemini Planet Imager on 2015 April 08 \citep[i.e. where the planet was at roughly the time when the ALMA observations were taken;][]{rameau16}.}
	\label{fcoint}
\end{figure}

Given the low signal to noise ratio here, it is worth checking to make sure that this is not a spurious line that appears in only one observation. This line does, indeed, appear in both data sets. In fact it is seen at $>3\sigma$ significance when considering just data set A, where the peak is 6.2$\pm$1.8~mJy at 8.3~\kms{}. In data set B, it shows up with a lower peak of 3.6$\pm$2.2~mJy at 9.5~\kms{}. Whilst the value for the peak in data set B is somewhat lower than that for data set A, they are consistent and the lower significance of the peak in data set B is also due to the worse observing conditions at the time of those observations \citep[see table 1 in ][]{su17a}.

It is also worth considering the false positive potential of our result. Our peak in the spectro-spatial filtered spectrum is significant at the 3.05$\sigma$ level. Given that we have 3791 channels (after removing flagged channels and those not covered by both data sets), we expect $3791\times0.5\times\erfc\left(\frac{3.05}{\sqrt{2}}\right)=4.3$ of them to have values $>+3.05\sigma$ even when no signal is present (given that the noise is approximately normally distributed). We, in fact, find that there are 4 channels with values $>+3.05\sigma$, thus making it plausible that this peak is simply noise. Nonetheless, the coincidence with (one measure of) the stellar radial velocity can give us some reassurance that the emission detected may be real. To calculate the false positive probability here we should consider the probability of a channel that has a velocity consistent with the stellar radial velocity (given the inconsistency of the two measurements of the stellar radial velocity noted above, we conservatively define this as within 2$\sigma$ of either of the measurements) and has a flux density $>3.05\sigma$ in either of the spectro-spatial filtered spectra. In other words, there are 80 channels in which the appearance of a line of this significance would be considered a detection. The probability of one of these containing noise of at least this level is roughly 8\%, i.e. low enough that we consider this a tentative detection but further observations are clearly required to determine for sure whether any CO is present in this system.

A Gaussian fit to the filtered spectrum with the East side moving towards us gives a peak in the emission of 4.6$\pm$1.1~mJy, at a velocity of \pvel{} and with a full width at half maximum (FWHM) of 1.9$\pm$0.5~\kms{}. The uncertainties on these values are calculated from the square root of the diagonal of the covariance matrix output by the curve fitting algorithm. Integrating over the line (which is found to cover 6 channels) gives an integrated line flux of \intlflux{} (where the uncertainty is calculated using equation \ref{eres} from the Appendix).

\begin{figure}
	\centering
	\includegraphics[width=0.48\textwidth]{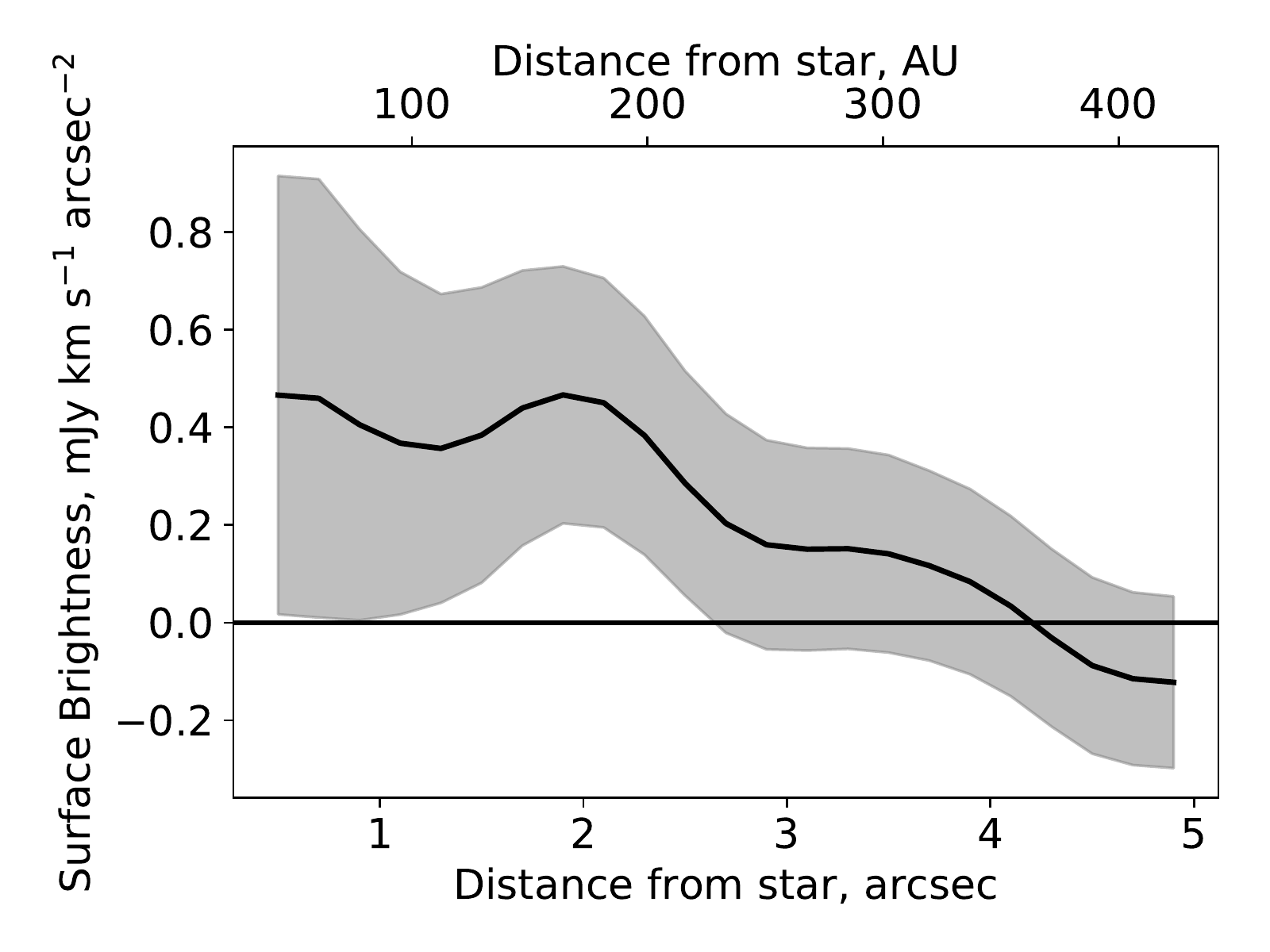}
	\includegraphics[width=0.48\textwidth]{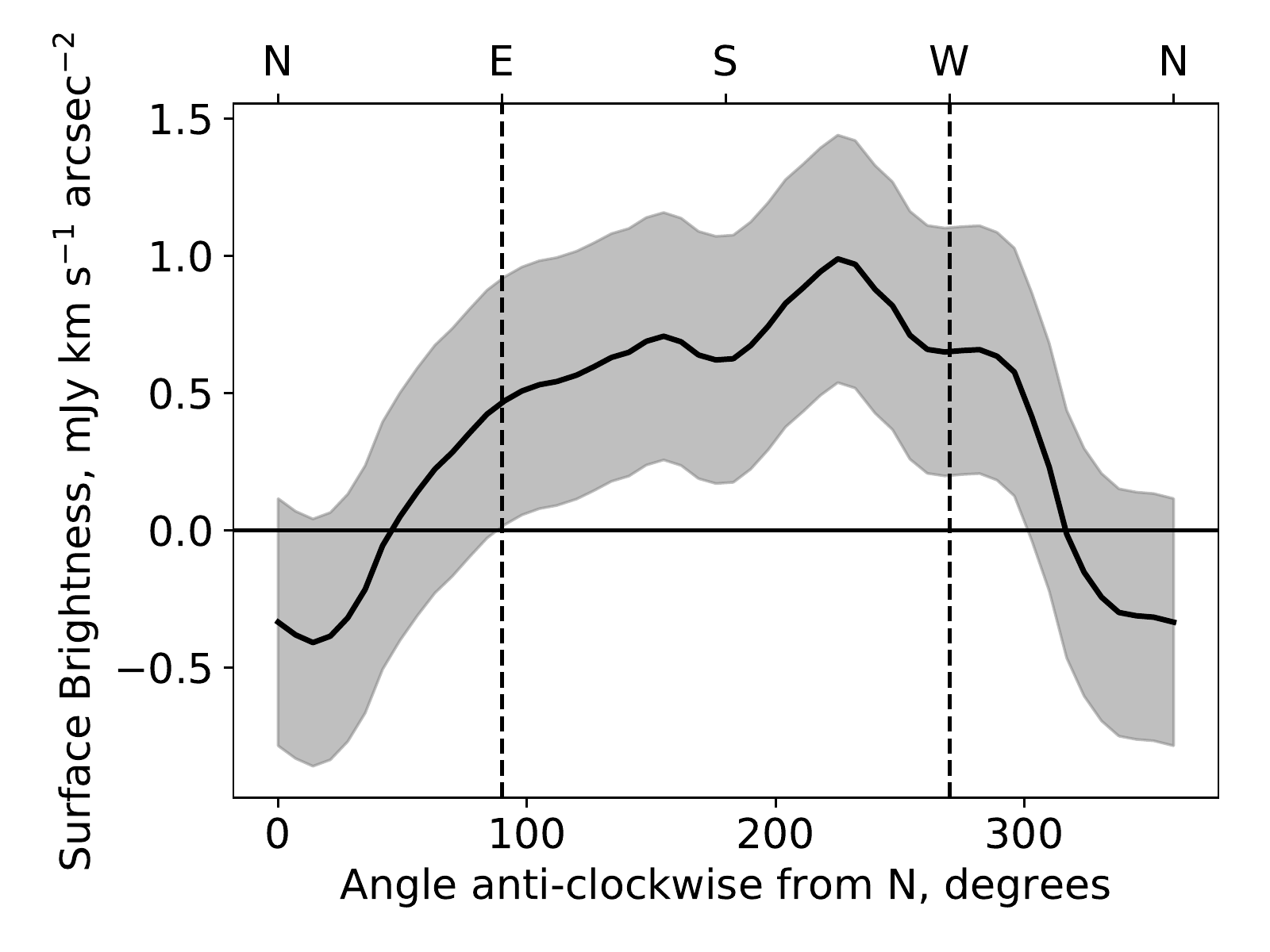}
	\includegraphics[width=0.48\textwidth]{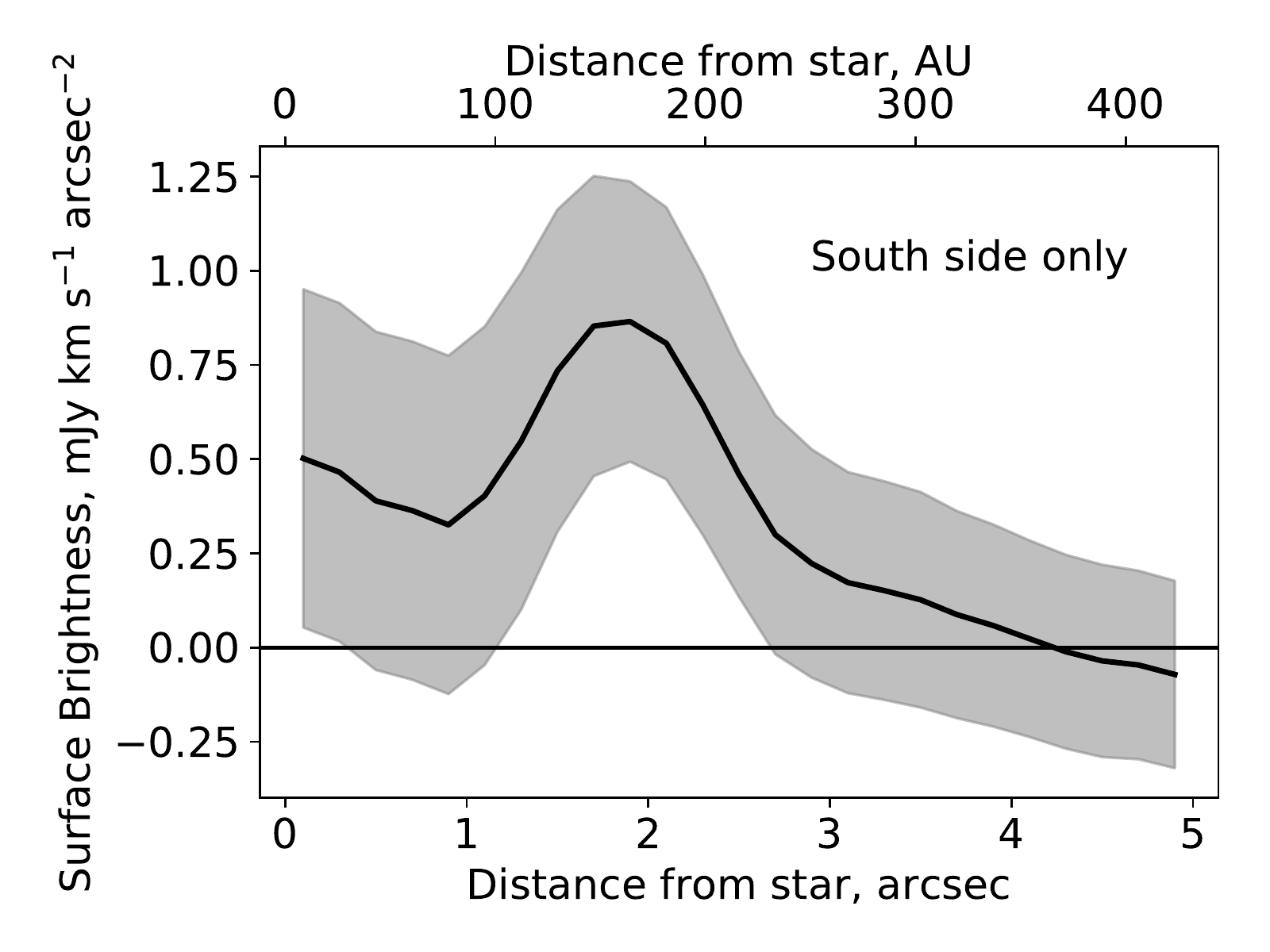}
	\caption{\emph{Top:} Deprojected radial distribution of CO $J$=2-1 line emission in the disc. This shows a running average such that each point is the weighted average within a 0.6\arcsec{} annulus. \emph{Middle:} Deprojected azimuthal distribution of gas in the disc. This shows a running average such that each point is the weighted average within a 45\degr{} wide sector extending from 100 to 220~AU. The vertical lines show the limits for the region referred to as the `South side' in the following and the main text. \emph{Bottom:} Similar to the top plot but only including emission on the South side of the disc (between $90\degr$ and $270\degr$) as the emission appears to be concentrated on one side. For each of the plots the grey area represents the 1$\sigma$ uncertainty, where $\sigma$ is calculated as described in the text.}
	\label{fcorad}
\end{figure}

Figure \ref{fcoint} shows the image obtained if we take this spectral shifted data cube and integrate over the six channels covered by the line. In figure \ref{fcorad} we show the radial and azimuthal distributions of the gas. In each case a running average is shown and the uncertainties shown are given by the standard error of the weighted mean\footnote{A weighted mean is used to account for the variation in sensitivity, which decreases away from the phase centre as a function of the primary beam level.} multiplied by the square root of the number of pixels per spatial beam to account for correlated noise. Although the noise is large, it is clear from these that the CO gas is concentrated on the South side of the disc (i.e. between $90\degr$ and $270\degr$) and is largely between 100 and 220~AU.

Given these results, we can now repeat the spectro-spatial filtering analysis but only including the South side of the disc. Fitting a Gaussian to this gives a peak in the emission of 3.3$\pm$0.7~mJy, centred at 8.4$\pm$0.2~\kms{} and with a FWHM of 2.2$\pm$0.5~\kms{}. The integrated line flux for this part of the disc is then 7.9$\pm$2.6~mJy\,\kms. This means that 83$\pm$42\% of the gas emission is coming from the South side of the disc. If this concentration of gas is real, it may be due to the concentration of planetesimals in a resonance with the planet, which is currently to the South-East of the star.

\subsection{CO $J$=3-2 analysis}
For the band 7 data we carry out a similar analysis to that in section \ref{s21}. Here we do not find any evidence of an emission line consistent with the velocity of that seen in the band 6 data even when the spectral filtering method is used. We find an integrated RMS of 34~\mjykms{} for a spatially integrated unresolved line. Given the short duration of the band 7 observations, we are only able to place a 3$\sigma$ upper limit of 190~\mjykms{} at the stellar velocity when assuming the same line width as the CO $J$=2-1 tentative detection. This implies a line ratio\footnote{Note that line ratios given in this paper are the ratios of the velocity integrated line fluxes. Care should be taken when comparing to other papers, as the ratio of the frequency integrated line fluxes is sometimes quoted.} for the disk of $F_{J\rm{=3-2}}/F_{J\rm{=2-1}}\leq20$, consistent with the expectations of models and measured excitation temperatures in other gas-bearing debris disks \citep[e.g.][]{hughes17,matra17}. We find that, as expected, this CO $J$=3-2 observation is not sensitive enough to achieve detection of the CO tentatively detected in the $J$=2-1 line. This lack of sensitivity to CO $J$=3-2 emission does not allow us to meaningfully constrain the excitation temperature or optical depth of the disc.

\begin{figure}
	\centering
	\includegraphics[width=0.48\textwidth]{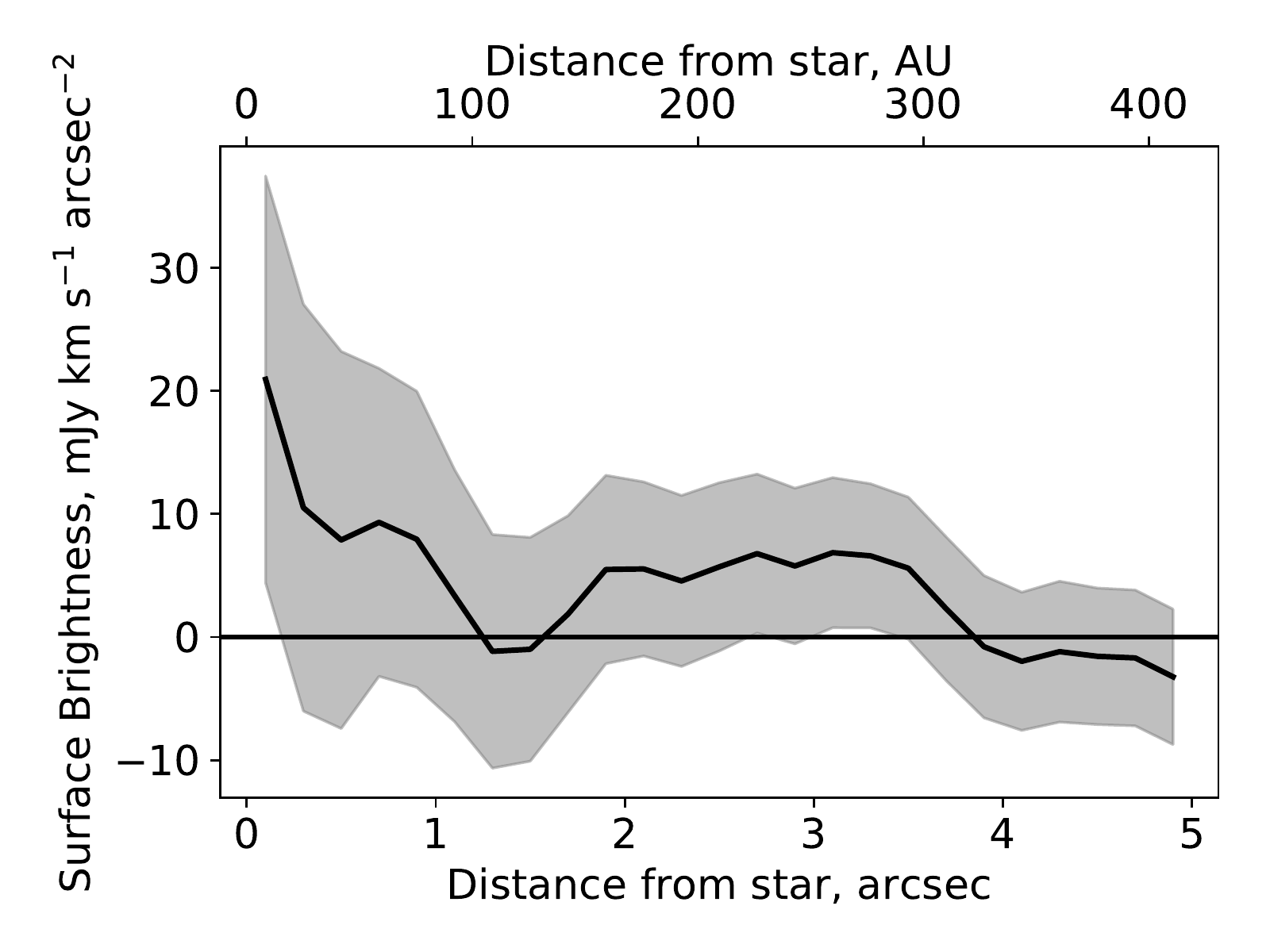}
	\caption{Deprojected radial distribution of the line detected in the band 7 data. The flat radial distribution indicates that this is likely to be noise (compare with figure \ref{fcorad}).}
	\label{fcoradb7}
\end{figure}

Intriguingly, when spatially integrating across the disc, we do find a significant line with a peak flux of 140$\pm$30~mJy, an integrated line flux of $200\pm60$~\mjykms{} and a FWHM of 1.3$\pm$0.3~\kms{} at a velocity of 2.3~\kms{}, i.e., inconsistent with both of the reported values of the stellar radial velocity. Given that this is a 4.6$\sigma$ detection, the chance of finding positive noise of such an amplitude in 3811 channels is just 0.7\%. Significantly, though, we find that spectral filtering actually reduces the significance of the detection no matter which orbit direction for the gas is assumed. Since no line is seen at this velocity in band 6 (for which the 3$\sigma$ upper limit on a line of the same width is 9.3~\mjykms{}), this means that the line ratio is $F_{J\rm{=3-2}}/F_{J\rm{=2-1}}\geq6.5$, which is higher than the line ratio of $\sim$0.3 expected for the fiducial model parameters of \citet{kral17}. We also find that if we integrate over the channels covered by the candidate line and azimuthally average, the radial profile is consistent with there being no disc i.e. there is no rise and fall as seen for the radial profile of the CO $J$=2-1 line (figure \ref{fcoradb7}). All this strongly suggests that this line is not associated with the disc and so is most likely just noise.

\subsection{CO $J$=1-0 analysis}
\label{s10}
\begin{figure}
	\centering
	\includegraphics[width=0.50\textwidth]{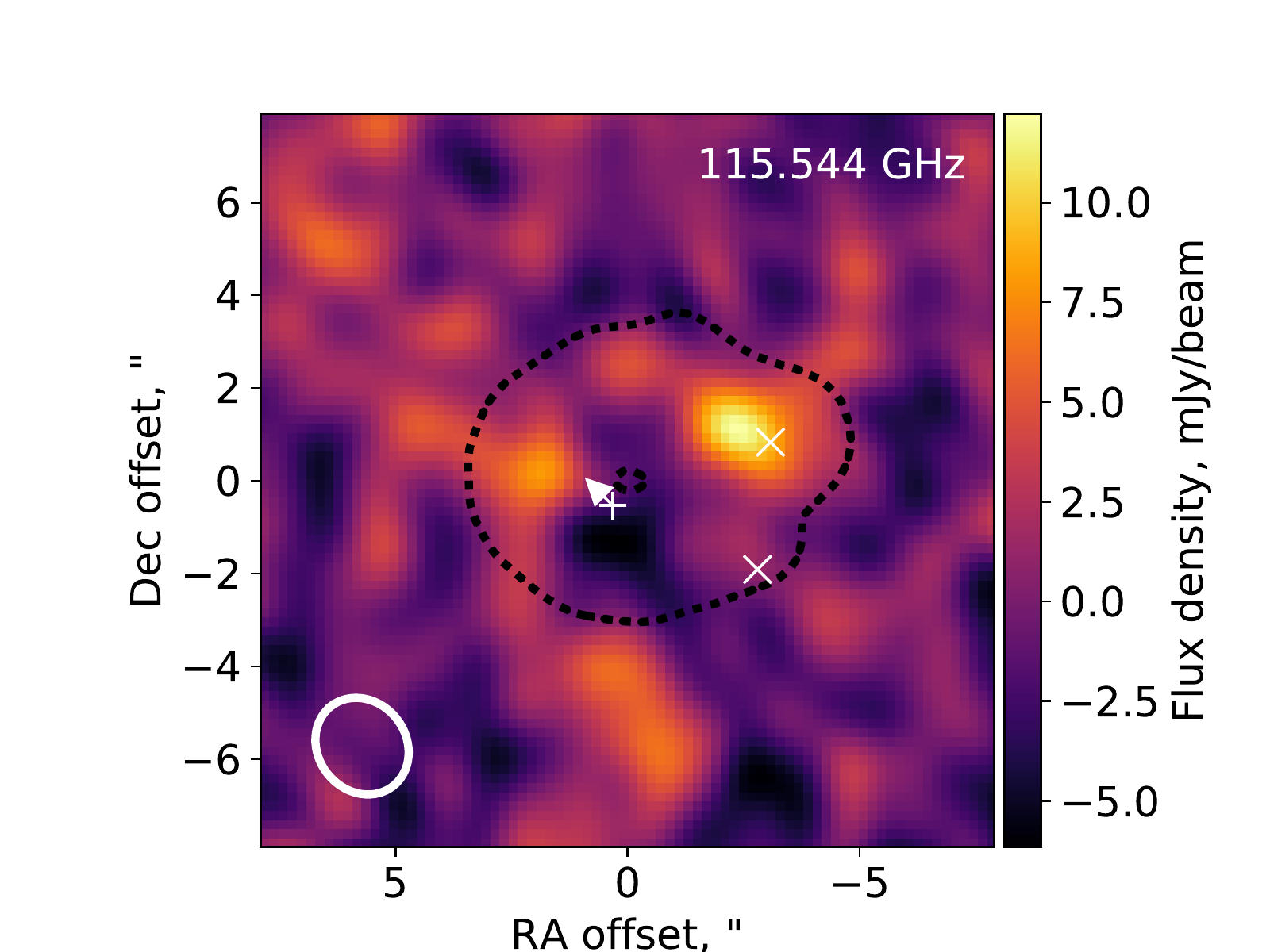}
	\caption{Image slice at the 115.544~GHz channel showing a point source consistent with the bright continuum source (the upper of the two white crosses).}
	\label{fb3im}
\end{figure}
The same analysis is also carried out on the band 3 data. In this case no significant lines show up in the spectrum when integrating over the disc, giving us an integrated RMS of 10~\mjykms{} for an unresolved line. The 3$\sigma$ upper limit of 46~\mjykms{} at the stellar velocity (assuming the same line width as the CO $J$=2-1 tentative detection) implies a line ratio for the disk of $F_{J\rm{=2-1}}/F_{J\rm{=1-0}}\geq0.2$, which, as with the $J$=3-2 analysis in the previous section, does not allow us to meaningfully constrain the excitation temperature or optical depth of the disc as the CO mass limit from this line ratio is always above the CO mass inferred from the CO $J$=2-1 analysis (see section \ref{sgas}).

When considering the bright continuum source, on the other hand, we do find a source at a frequency of 115.544$\pm$0.002~GHz that is both spatially and spectrally unresolved. It has a flux density of 12.2$\pm$2.2~mJy and, given the beam size, is consistent with the location of the bright continuum source as shown in figure \ref{fb3im}. However, since this is unresolved spectrally, it cannot be coming from a background galaxy as galaxies have CO linewidths of $>$100~\kms{}. The frequency means it would have a radial velocity of $-712$~\kms{} if it is CO $J$=1-0, which is unusually high for a galactic background source. We therefore conclude that this is a spurious source of unknown origin unrelated to the continuum emission.

\section{Resolved spectral index map}
\label{sspec}
\begin{figure*}
	\centering
	\includegraphics[width=0.45\textwidth]{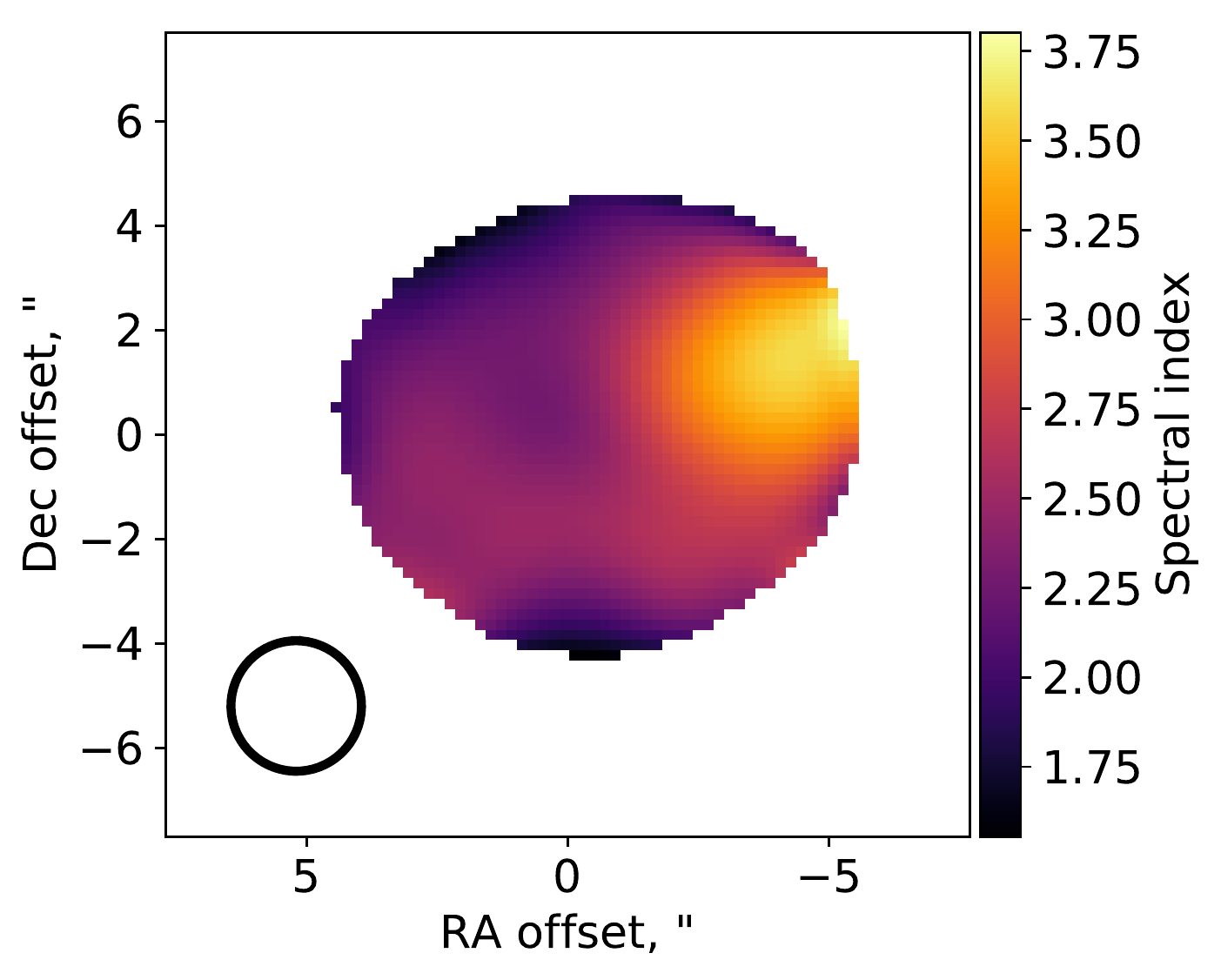}
	\includegraphics[width=0.51\textwidth]{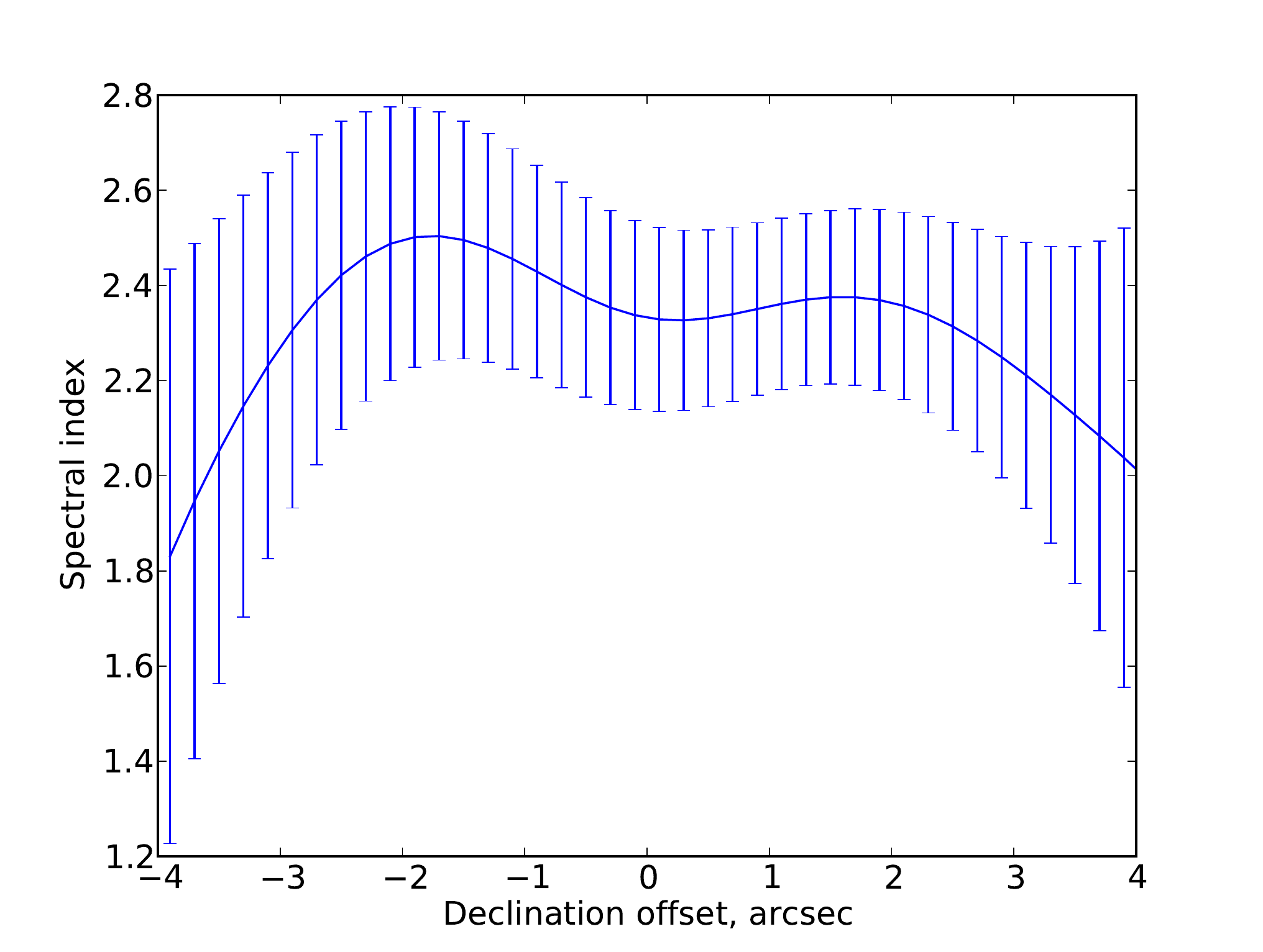}
	\caption{\emph{Left:} Spectral index map created by fitting power laws to the band 3, 6 and 7 images (for details see section \ref{sspec}). The bright source clearly has a much steeper spectral index than the disc. \emph{Right:} Line-cut from South to North along the centre of the disc with 1$\sigma$ error bars calculated from the power law fit.}
	\label{fspec}
\end{figure*}

In addition to probing multiple gas lines, observing the disc at multiple ALMA wavelengths allows us to determine the spectral index of the dust across the image. At submillimetre wavelengths (i.e. in the Rayleigh-Jeans limit), the flux density, $F_{\nu}$, is a power law function of the wavelength $\lambda$ in the form $F_{\nu}\propto\lambda^{-\alpha}$, where $\alpha$ is the spectral index. If the grains emit as pure blackbodies, they would have a spectral index of 2. In reality, dust grains are inefficient emitters at wavelengths much greater than their size and so we expect deviations from this at sub-millimetre wavelengths. To determine the spectral index we create CLEAN, naturally weighted images of the continuum emission for each pair of spectral windows. This gives us seven images (two in band 3, three in band 6 due to the overlap between datasets A and B and two in band 7) at frequencies of 102, 114, 216, 231, 246, 333 and 345~GHz. We smooth them all to a consistent 2.5\arcsec{} resolution (chosen to be slightly larger than the beam size of the band 3 image). Then for each pixel we fit a power law to get the spectral index and its uncertainty (which is calculated from the square root of the diagonal of the covariance matrix output by the curve fitting algorithm). The resultant spectral indices for the continuum emission are shown (only spectral indices with an SNR$>$2 in order to only show spectral indices for the real emission) in figure \ref{fspec} (left). We find that the bright source clearly has a much steeper spectral index than the disc, with a peak at 3.6$\pm$0.3 (which will be discussed further in section \ref{sbright}). Some variation in the spectral index is seen across the disc but this is not found to be significant (see figure \ref{fspec} right). A weighted average across the disc (defined here to be pixels with a spectral index that has an SNR$>$2 and flux density in the smoothed band 7 image of less than 60~$\mu$Jy\,beam$^{-1}$ to avoid the bright source) results in a spectral index of 2.41$\pm$0.12 (which will be discussed further in section \ref{ssize}).

\section{Discussion}
\subsection{Gas production within the disc}
\label{sgas}
In the previous cases where gas has been detected in debris discs, there are some discs where the gas is expected to be primordial gas and others where the gas must be second generation. As noted by \citet{zapata18}, the low gas mass in this system clearly rules out the presence of primordial gas. In this section we check whether it is consistent with current models for second generation production through collisional cascades of icy planetesimals and make predictions for the possibility of confirming the presence of CO by searching for other emission lines.

In section \ref{s21}, we found an integrated CO $J$=2-1 line flux of \intlflux. Using the molecular extinction code of \citet{matra15} \citep[enhanced to include UV fluorescence as discussed in ][]{matra18} we can convert this flux to a CO mass. Due to its steep increase as a function of UV wavelength, the stellar UV dominates over the interstellar UV at the wavelengths relevant for CO fluorescence. The UV field was, therefore, modelled using a PHOENIX stellar atmosphere model (Arkenberg et al. in prep.)\footnote{The latest PHOENIX models can be found at \url{https://www.astro.uni-jena.de/Users/theory/for2285-phoenix/grid.php}.} with the known stellar parameters of HD95086, and scaled to match the star's flux as seen from Earth. The conversion also depends on the density of collisional partners (assumed to be electrons for a second generation gas production scenario) and the kinetic temperature of the gas. These quantities are unknown, but a range of values can be used that is sufficiently wide to cover the transition between the two limiting regimes of radiation-dominated at low electron density and collision-dominated at high electron density (specifically, we use electron densities between 10$^{-4}$ and 10$^{10}$~cm$^{-3}$ and kinetic temperatures between 10 and 250~K). We, therefore, find that our integrated line flux translates to a CO mass of \comass. 

Undertaking the same analysis using the CO $J$=3-2 and $J$=1-0 flux limits results in limits on the CO mass that are always above the mass found from the $J$=2-1 emission line and so these non-detections do not help constrain the excitation temperature or electron density.
Conversely, we can use the CO mass calculated from the CO $J$=2-1 observations to make predictions for the sensitivity needed to detect these other emission lines. These predictions are dependent on the line ratio, which is dependent on the assumed electron density and kinetic temperature. Using the same ranges as before, we predict integrated line fluxes for CO $J$=3-2 between 2 and 40\mjykms{} and CO $J$=1-0 between 0.6 and 10\mjykms{}, where low electron densities in the radiation-dominated regime result in the highest CO $J$=1-0 line flux and lowest CO $J$=3-2 line flux whereas high electron densities in the collision-dominated regime produce the opposite. Assuming the same set-up as used for the band 7 and band 3 observations presented here (i.e. multiplying the observation time of the observations presented here by the square of the required increase in sensitivity), a 3$\sigma$ detection of the CO $J$=3-2 line would take between 40 minutes and 10 days, whilst a 3$\sigma$ detection of the CO $J$=1-0 line would take between 8 hours and 2 years of observation time (not including calibration time). In other words, it may be possible to detect one of the other CO emission lines within a few hours, although the optimum choice depends on the unknown line ratio and completely rejecting the presence of CO at a level consistent with our tentative detection would require unfeasibly long observations with ALMA.

\citet{kral17} predicted line flux densities for a large sample of systems based on a model of CO gas produced in a steady state collisional cascade. For previous gas detections and non-detections they found their model fitted the observations to within an order of magnitude except in a couple of systems where the gas appears to be produced in a different manner. HD~95086 is part of their sample, although for improved accuracy we re-do their calculations using new data from \emph{Gaia} for the distance and luminosity, and using the fit to the ALMA continuum data \citep[$R_0=200$~AU, $dr/r=1$;][]{su17a}. This results in a predicted CO mass of $\sim6\times10^{-6}$~M$_\oplus$. Given that this lies within the range found by our observations, we conclude that our observations are consistent with this model of gas produced through the collisional cascade. Using our CO mass estimate and our expectation that the CO is produced from a collisional cascade, we can then determine the  CO+CO$_2$ ice mass fraction. This is calculated using equation 2 from \citet{matra17a}:
\begin{equation}
 f_{\rm{CO+CO_2}}=\frac{1}{1+0.0012R^{1.5}\Delta R^{-1}f^2L_\star M_\star^{-0.5}t_{\rm{phd}}M_{\rm{CO}}^{-1}},
 \label{efco}
\end{equation}
where $R$ and $\Delta R$ are in AU and are the radius and width of the belt respectively, $f$ is the fractional luminosity, $L_\star$ is the luminosity of the star and $t_{\rm{phd}}$ is the timescale for photodissociation \citep[which is $\sim$120 years if we take the interstellar radiation field as the dominant source of photodissociating radiation, as is the case for Fomalhaut; ][]{matra15}. This results in a CO+CO$_2$ mass fraction in the exocomets of 5.7-35.7\%, which is consistent with the few other belts where this has so far been measured and with direct inheritance of ISM compositions \citep{matra17a}.

We also note that as CO is photodissociated into carbon and oxygen, ionised carbon and oxygen will build up in the system, spreading inwards towards the star. With ALMA it is possible to observe \ion{C}{I} emission in band 8 or 10 and we expect this to have a higher line flux than the CO emission since \ion{C}{I} is not so quickly removed from the system and so the mass is much higher \citep{kral17}. With the right combination of sensitivity and resolution, it could even be possible to detect a gap in \ion{C}{I} caused by the planet at $\sim$54~AU. Alternatively, if the CO is produced from a recent collision or collisions, then the \ion{C}{I} could also be concentrated, as is seen in the case of $\beta$ Pic \citep{cataldi18}.

\subsection{Geometry of the disc}
In section \ref{s21} we found that the $J$=2-1 signal was increased through spectro-spatial filtering when we assumed that the gas is moving towards us from the East ansa and away at the West ansa. As \citet{matra17a} note, this information alone cannot tell us which side of the disc is inclined towards us and which is inclined away from us as the CO cannot tell us whether the disc is rotating clockwise or anti-clockwise. However, there have been enough observations of the planet in the system to show that it is orbiting in a clockwise manner \citep{rameau16}. Since the planet and the debris disc will have formed out of the same protoplanetary disc, we can assume that the particles in the disc are also orbiting in a clockwise direction. By combining this on sky rotation with the radial information from the CO, we find that the northern side of the disc is the near side. 

An alternative way to determine the near side of a debris disc is by looking at asymmetries in the scattered light. For instance, HR 4796 shows a strong asymmetry with the western side of the disc being much brighter than the eastern side in polarised scattered light, which is interpreted as meaning the western side is nearer \citep{milli15,perrin15}. The debris disc around HD~95086 has also recently been detected in polarised scattered light \citep{chauvin18}. Unfortunately, as the disc is wide and close to face-on, the surface brightness is low and so the detection can only be made by azimuthally averaging. Therefore, higher sensitivity scattered light observations are necessary to  determine if there is any brightness asymmetry between the two sides of the disc. Even then, it is not always clear from scattered light observations which side is nearer as it is dependent on whether the dust is forward-scattering or back-scattering. Most of the few cases where observations of the scattering phase function in debris discs and cometary dust in the Solar System show the dust to be forward-scattering \citep[see][and references therein]{hughes18}. However, there may be at least one exception to this rule: Fomalhaut. In this case, both the rotation of the star \mbox{\citep{lebouquin09}} and the orbital motion of the gas \citep{matra17a} point towards the brighter side being further away from us assuming Fomalhaut b is on a prograde orbit. This implies that the grains would be back-scattering \citep{min10} in the Fomalhaut disc. This demonstrates the benefit of combining gas, scattered light and planet observations to determine the correct geometry and orbital direction of a planetary system.

\subsection{Nature of the compact continuum sources}
\label{sbright}
In addition to the disc, the ALMA continuum image shows two compact sources coincident with the West side of the disc; one bright, marginally resolved source and one faint, unresolved source \citep{su17a}. In this section we discuss the possible nature of these sources. 

\citet{zapata18} used the proper motion of the star to provide further evidence that the bright source must be a background source. The lack of detection of the star meant that they had to infer the position by modelling the disc (for which they used a trial and error method) and assuming the star is at the centre. They find that the centre of the ring is moving with a proper motion (in right ascension and declination) of $\mu_\alpha=-90\pm25$~mas\,yr$^{-1}$ and $\mu_\delta=-31\pm27$~mas\,yr$^{-1}$ compared to the \emph{Gaia} DR2 values of $\mu_\alpha=-41.14\pm0.05$~mas\,yr$^{-1}$ and $\mu_\delta=12.70\pm0.05$~mas\,yr$^{-1}$ \citep{gaia18}, whereas the bright source remains in roughly the same location. Given the large uncertainties and the assumptions in the method to calculate them, we consider this evidence inconclusive. The low proper motion of this source means that it will likely be necessary to wait many more years before a proper motion difference can conclusively be determined and, unfortunately, if the bright source is a background object, it will still contaminate the observations for the next $\sim$200 years as the star is moving towards the current location of the bright source.

\citet{su17a} considered a number of possibilities for what the bright source could be if it was part of the disc. In all cases, CO gas is expected to be concentrated at the same location as the enhanced dust emission, as is seen in $\beta$ Pic \citep{dent14}. We confirm here that there is no significant concentration of CO gas present at the locations of these sources, even when taking into account the Keplerian velocities (figure \ref{fcoint}). Specifically for the band 6 observations (the deepest and most constraining observations discussed in this paper), the RMS in a single channel is 0.24~\mjykmsbm{} and the area of the bright source is roughly 1.4 beams so that the 3$\sigma$ upper limit of the line flux is $\sim$1~mJy\,\kms{}. This is $<11\%$ of the total line flux density i.e. the fraction of CO gas in this region compared to the whole disc is much less than the fraction of continuum emission in this region compared to the whole disc ($\sim25\%$). Here we have assumed that the range of velocities in the clump is low such that the line is unresolved, although we note that it is possible that the range of velocities is high enough that the line becomes resolved, thereby marginally increasing the upper limit. Nonetheless, it seems unlikely that the bright continuum source is contributing any CO around the stellar velocity.

\citet{su17a} also noted the steeper spectral index of the bright source compared to the disc using just the band 6 data. In section \ref{sspec} we make use of the new band 3 and 7 data to give us a much more accurate measure of the spectral index and confirm that the spectral index of the bright source is indeed steep with a value at the peak of 3.6$\pm$0.3 compared to 2.41$\pm$0.12 for the rest of the disc (as discussed in section \ref{ssize}). As discussed in \citet{su17a}, this steep spectral index is consistent with this source being a background galaxy. For instance, \citet{casey12} found from a sample of luminous and ultraluminous infrared galaxies, that the spectral slope in the Rayleigh-Jeans regime is 3.60$\pm$0.38.

In conclusion, the lack of a concentration of CO at the location of the bright source and the clear difference in the spectral index between the source and the disc shown here does back-up the earlier claims of this source being an unrelated background source.

For the faint, unresolved source (indicated by the lower cross in figure \ref{fcoint}), the large uncertainties here mean that we cannot draw any conclusions as to whether that point contributes strongly to the CO emission or not and so the possibility remains open that this is a clump at the outer edge of the disc.

\subsection{Size distribution}
\label{ssize}
The spectral energy distributions of debris discs are close to blackbodies but deviate from the Rayleigh-Jeans slope at long wavelengths since dust grains are inefficient emitters at wavelengths much greater than their size. Measuring the spectral slope at submillimetre wavelengths can, therefore, allow us to determine the size distribution of grains in the disc. By using resolved images we can look for variations in the spectral index across the disc and avoid contamination from the background source (see section \ref{sspec}). By doing this we find that there are no significant variations in spectral slope across the disc and find an average of 2.41$\pm$0.12. This is consistent with the value of 2.37$\pm$0.15 found by \citet{macgregor16} using the ALMA 1.3~mm photometry \citep{su17a} and the ATCA 7~mm photometry \citep{ricci15}. By using the analytic approximations of \citet{draine06}, which assume an astrosilicate composition, we can interpret our measure of the spectral index as  a size distribution slope of 3.27$\pm$0.07. This is quite shallow compared to spectral indices of other debris discs, but not without precedent \citep{macgregor16,holland17}.

Whilst the variations in spectral index across the disc seen in figure \ref{fspec} are not found to be significant due to the low sensitivity of the band 3 and 7 data and the low resolution of the band 3 data, with deep resolved maps at multiple ALMA wavelengths, such methods could be used to determine such variations. Variations are expected, for instance, in the case of an eccentric disc, where a difference in size distribution is expected between the pericentre and apocentre due to it being easier to remove small grains by radiation pressure at the pericentre than at the apocentre \citep{lohne17,kim18}.

\section{Conclusions}
In this paper we search for CO gas in the debris disc around the star HD 95086 using ALMA. We find a tentative signal for the $J$=2-1 line. This candidate line has a peak of 4.3$\pm$1.4~mJy at a velocity of \pvel. This velocity is consistent with the stellar radial velocity found by \citet{madsen02}, the significance of the line is higher when spectro-spatially filtering compared to just spatially filtering and there is a clear rise in the radial profile consistent with the location of the belt as seen in continuum emission. We therefore consider this to be a tentative detection of CO gas in the HD 95086 system, although it is clear that more observations are needed to confirm this. We find that the data are best matched by gas moving towards us from the East ansa and away at the West ansa. Assuming the gas is orbiting in the same direction as the known planet (clockwise), then we can say that the South side of the disc and the planet are inclined away from us and the North side is the near side. The integrated line flux found is \intlflux, corresponding to a CO mass of \comass{}, which would imply a cometary mass fraction of 5.7-35.7\%. This is consistent with second generation production through the collisional cascade as derived by \citet{kral17}.

We did not find any evidence for CO $J$=3-2 or $J$=1-0 emission. This is unsurprising given the much lower sensitivity of the band 3 and band 7 datasets and so does not give us any strong constraints on the CO line ratios. Whilst it is hard to make predictions for how deep we would need to go to detect the CO emission at these bands as we do not know for sure what the line ratio is, we predict that CO could be detected in either band 7 or band 3 observations within a few hours depending on the excitation regime. As noted by \citet{kral17}, looking for \ion{C}{I} in band 8 would also be a viable and, perhaps, more promising strategy to follow-up this tentative detection.

Contrary to the continuum observations, we find no bright point-like CO source consistent with the radial velocity of the system. We make use of the continuum images in all three bands to precisely determine the spectral index across the image and find a clear difference between the bright source and the disc. Both of these factors add weight to the hypothesis that the continuum source is an unrelated background object.

\section*{Acknowledgements}
The authors thank Steve Ertel, Sarah Morrison and Renu Malhotra for providing comments on a draft of the manuscript and also thank the referee for a thorough and constructive review. MB acknowledges support from the Deutsche Forschungsgemeinschaft through project Kr 2164/15-1. LM acknowledges support from the Smithsonian Institution as a Submillimeter Array (SMA) Fellow. QK acknowledges support from STFC via the Institute of Astronomy, Cambridge Consolidated Grant. AMH acknowledges support from National Science Foundation (NSF) grant AST-1412647. MAM acknowledges support from a NSF Astronomy and Astrophysics Postdoctoral Fellowship under Award No. AST-1701406. TL acknowledges support from the Deutsche Forschungsgemeinschaft through project Lo 1715/2-1. This paper makes use of the following ALMA data: \\ ADS/JAO.ALMA\#2013.1.00773.S, \\ ADS/JAO.ALMA\#2013.1.00612.S, \\ ADS/JAO.ALMA\#2016.A.00021.T. \\ ALMA is a partnership of ESO (representing its member states), NSF (USA) and NINS (Japan), together with NRC (Canada) and NSC and ASIAA (Taiwan) and KASI (Republic of Korea), in cooperation with the Republic of Chile. The Joint ALMA Observatory is operated by ESO, AUI/NRAO and NAOJ. The National Radio Astronomy Observatory is a facility of the National Science Foundation operated under cooperative agreement by Associated Universities, Inc. This research made use of Astropy, a community-developed core Python package for Astronomy \citep{astropy13}.

\bibliographystyle{mnras}
\bibliography{thesis}{}

\appendix
\section{Uncertainty on the integrated line flux}
\label{sapp1}
The RMS on the integrated line flux for an unresolved line, $\sigma_{\rm{int}}$, is given by
\begin{equation}
\sigma_{\rm{int}}=\sigma_{\rm{sp}} w_{\rm{e}}, 
\label{eunres}
\end{equation}
where $\sigma_{\rm{sp}}$ is the RMS of the integrated spectrum and $w_{\rm{e}}$ is the effective bandwidth, which is the channel width, $w_{\rm{ch}}$, multiplied by the number of channels in the effective bandwidth, $b_{\rm{e}}$ (2.667 in this case as Hanning smoothing was used). This takes into account that neighbouring channels are correlated, so the RMS measured in the spectrum is not characteristic of one channel, but of one effective bandwidth\footnote{\url{https://safe.nrao.edu/wiki/pub/Main/ALMAWindowFunctions/Note_on_Spectral_Response.pdf}}.

When the line is resolved, i.e. the number of channels, $N_{\rm{ch}}>b_{\rm{e}}$, then this is changed to 
\begin{equation}
\sigma_{\rm{int}}=\sqrt{(\sigma_{\rm{sp}} w_{\rm{e}}\sqrt{N_{\rm{ich}}})^2+\sigma_{\rm{cal}}^2},
\label{eres}
\end{equation}
where $N_{\rm{ich}}$ is the number of independent channels, i.e. $N_{\rm{ich}}=N_{\rm{ch}}/b_{\rm{e}}$. For completeness, we have also added in quadrature the flux calibration uncertainty, $\sigma_{\rm{cal}}$ \citep[assumed to be 5\% of the integrated flux; see section C.4.1][]{alma16}.

\bsp

\end{document}